\documentclass[11pt]{article}
\usepackage{graphicx}
\usepackage[margin=1.25in]{geometry}
\usepackage[usenames,dvipsnames]{color}
\usepackage{url}
\usepackage[colorlinks = true,
            linkcolor = blue,
            urlcolor  = blue,
            citecolor = blue,
            anchorcolor = blue]{hyperref}

\newcommand\pubnumber{Transcendental Preprint }
\newcommand\pubdate{\today}

\def\Title#1{\begin{center} {\LARGE #1 } \end{center}}
\def\Author#1{\begin{center}{ \sc #1} \end{center}}
\def\Address#1{\begin{center}{ \it #1} \end{center}}

\newcommand\pubblock{\rightline{\begin{tabular}{l} \pubnumber\\
         \pubdate \end{tabular}}}
\newenvironment{Abstract}{\begin{quotation} \begin{center}
                       ABSTRACT
     \end{center}\bigskip  }{\end{quotation}}

\def\beq{\begin{equation}}
\def\eeq#1{\label{#1}\end{equation}}
\def\eeqn{\end{equation}}

\newenvironment{Eqnarray}%
   {\arraycolsep 0.14em\begin{eqnarray}}{\end{eqnarray}}
\def\beqa{\begin{Eqnarray}}
\def\eeqa#1{\label{#1}\end{Eqnarray}}
\def\eeqan{\end{Eqnarray}}

\let\bar=\overbar

\def\lsim{\mathrel{\raise.3ex\hbox{$<$\kern-.75em\lower1ex\hbox{$\sim$}}}}
\def\gsim{\mathrel{\raise.3ex\hbox{$>$\kern-.75em\lower1ex\hbox{$\sim$}}}}

\def\del{\partial}
\def\Dslash{\not{\hbox{\kern-4pt $D$}}}
\def\dslash{\not{\hbox{\kern-2pt $\del$}}}
\def\pslash{\not{\hbox{\kern-2pt $p$}}}
\def\ETmiss{\not{\hbox{\kern-4pt $E$}}_T}

\def\Dlr{\mathrel{\raise1.5ex\hbox{$\leftrightarrow$\kern-1em\lower1.5ex\hbox{$D$}}}}

\def\MSB{{\bar{M \kern -2pt S}}}
\def\msb{{\bar{\scriptsize M \kern -1pt S}}}

\def\drb{{\bar{\scriptsize D \kern -1pt R}}}

\newcommand\snowmass{\begin{center}\rule[-0.2in]{\hsize}{0.01in}\\\rule{\hsize}{0.01in}\\
\vskip 0.1in Submitted to the  Proceedings of the US Community Study\\ 
on the Future of Particle Physics (Snowmass 2021)\\ 
\rule{\hsize}{0.01in}\\\rule[+0.2in]{\hsize}{0.01in} \end{center}}

\begin{document}

\pubblock

\Title{Two-fermion final states at International Linear Collider}

\bigskip 

\Author{Taikan Suehara}

\medskip

\Address{ Department of Physics, Kyushu University, 744 Motooka, Nishi-ku, Fukuoka 819-0395, Japan}

\medskip

 \begin{Abstract}
\noindent The pair productions of leptons and quarks at $e^+e^-$ Higgs factory are an important probe for new physics via precise measurements. The discovery and exclusion limits of $Z^\prime$ models at the International Linear Collider with $\sqrt{s} = 250$, 500 and 1000 GeV are calculated with selection efficiencies estimated in the existing full simulation studies.
It shows a large potential of BSM searches with precise measurements at future energy-frontier $e^+e^-$ colliders.
\end{Abstract}

\snowmass

\def\thefootnote{\fnsymbol{footnote}}
\setcounter{footnote}{0}
%


\section{Introduction}

An electron-positron Higgs factory is one of the most promising next-generation energy-frontier collider projects in high energy physics. There are several plans to realize such a Higgs factory, based on a scheme of either circular or linear accelerator. Projects with circular accelerators, such as FCCee or CEPC have more luminosity on $\sqrt{s} = 250$ GeV or smaller, while linear accelerators, such as ILC or CLIC have possibility for energy upgrade and polarization of initial particles. The main purpose of the Higgs factory is to precisely measure parameters of Higgs bosons including coupling to standard model particles, but it can also be utilized as direct and indirect discovery of new particles. The 2-fermion final states, $e^+e^- \to f\bar{f}$, give strong discovery potential of BSM indirectly, thanks to the simple final states with very precise theoretical calculation possible and the relatively big cross sections.

In this study, ILC, or International Linear Collider, is used as a benchmark project of analyzing such final states. Since ILC is a linear machine, we can assume energy upgrades to 500 GeV and 1 TeV, which are expected to have higher sensitivity to the contribution of heavy new particles. We evaluate the discovery and exclusion potential of high energy machines with extrapolation of existing simulation studies in the lower center-of-mass energies. Another advantage of the ILC is the polarized beams. We utilize the dependence of the cross sections on the initial polarization as an additional observable. $P(e^-, e^+) = (\pm 0.8, \mp 0.3)$ are assumed for $\sqrt{s} = 250$ and 500 GeV, while $P(e^-, e^+) = (\pm 0.8, \mp 0.2)$ are assumed for $\sqrt{s} = 1$ TeV in this study. Integrated luminosities of 0.9, 1.6 and 3.2 ab$^{-1}$ are assumed for each polarization described above at $\sqrt{s} = 250$, 500, and 1000 GeV, respectively. The physics cases of the ILC, including 2-fermion final states in more comprehensive way, are reviewed in \cite{LCCPhysicsWorkingGroup:2019fvj}.

\section{A review: efficiency of $e^+e^- \to f\bar{f}$ final states at ILC}

There are several studies of $e^+e^- \to f\bar{f}$ using full detector simulation of ILD (International Large Detector, one of two validated ILC detector concepts). Efficiency of event selections of each final state of $ee$, $\mu\mu$, $\tau\tau$, $bb$ and $cc$ is reviewed in this section.

$e^+e^- \to e^+e^-$ and $\mu^+\mu^-$ are the simplest final states among them and thus high efficiency is expected. \cite{Deguchi:2019tvp} shows such a study with $\sqrt{s} = 250$ GeV ILC. The cut tables show a few to 10\% overall efficiency of the signal final states with much smaller remaining background. 
The relatively low signal efficiency is mainly caused by the mixture of signal with radiative events.
This is confirmed by checking signal efficiency with preselections of energy sum of two leptons in MC truth $> 237.5$ GeV and polar angle $|\cos\theta|$ of both leptons $< 0.95$.
The efficiency of the cuts with the preselected signal events are more than 98\% with $\mu$-pair events and more than 97\% with $e$-pair events.

Efficiency of $e^+e^- \to \tau^+\tau^-$ is estimated at \cite{Jeans:2019brt} with $\sqrt{s} = 500$ GeV.
In this study, $\tau$-pair with at least one hadronic decay is selected to reduce leptonic background. Overall efficiency with $\ge 1$ hadronic decay and $m_{\tau\tau} > 480$ GeV is about 60\%, but omitting last 2 cuts raises the efficiency by about 10\%.
Since the cuts are optimized for tau polarization measurement in the study, we take the loose cuts with 70\% efficiency. The remaining background is still much smaller than the signal events. Assuming the same efficiency for $\tau\tau \to e\nu\nu \mu\nu\nu$ and removing all other fully leptonic tau decays, the expected overall efficiency is as high as 65\%. 

$e^+e^- \to bb$ and $cc$ can also be used to probe the BSM contribution.
Although the production cross section is higher due to the color multiplicity, 
the selection is much more difficult in those final states with 2 heavy jets.
Since we use angular distribution, identification of the quark charge is required.
Based on \cite{Bilokin:2017lco}, we expect the efficiency of $bb$ and $cc$ final states as
29\% and 7\%, respectively.

In summary, we expect the efficiency of $e^+e^- \to f\bar{f}$ as:
\begin{itemize}
\item $ee$: 0.97,
\item $\mu\mu$: 0.98,
\item $\tau\tau$: 0.65,
\item $bb$: 0.29,
\item $cc$: 0.07,
\end{itemize}
which are used in the following discussions.

\section{Limits for $Z^\prime$ models}

Here we investigated possibility to search for $Z'$ models\cite{Hewett:1988xc}, where $Z'$ is an additional neutral vector gauge boson coupled to SM fermions.
The coupling constants differ depending on models, and we used SSM (Sequential Standard Model), and $E_6$ models. The SSM assumes the same coupling constants as SM $Z$.
On the other hand, the $E_6$ is a string-motivated model which naturally introduces $Z'$ as a linear combination of the two extra $U(1)$ gauge bosons $Z_{\psi}$ and $Z_{\chi}$ : $Z' =Z_{\chi}\cos\beta + Z_{\psi}\sin\beta$. We investigated three $\beta$ parameters: $\beta = 0$ ($\chi$ model), $\beta = \pi /2 $ ($\psi$ model) and $ \beta =\pi - \mathrm{arctan} \sqrt{5/3}$ ($\eta$ model).
ALR (Alternative Left-Right symmetric) is another model also introduced from $E_6$, which gives extra $SU(2)_R$ in addition to SM $SU(2)_L$. This introduces an additional $Z_R$ boson phenomenologically treated as $Z'$, which behaves like SM $Z$, but gives different couplings to SM particles.

The discovery potential and exclusion limits to each $Z^\prime$ model are calculated based on the tree-level differential cross sections on polar angles. The expected number of SM events are estimated in 20 bins of $\cos\theta$, from $-0.95$ to $0.95$, based on the cross sections, integrated luminosities written in the introduction, and signal efficiencies shown in the previous section. Systematic uncertainties of 0.1\% for $e$ and $\mu$ pairs, 0.2\% for $\tau$ and $b$ pairs, and 0.5\% for $c$ pairs are assumed at each bin with random distributions.
Deviations caused by $Z^\prime$ models are used to calculate separation power with changing $Z^\prime$ masses. The mass limits of 95\% exclusion and $5\sigma$ discovery are calculated at $\sqrt{s} = 250$, 500 and 1000 GeV separately. The obtained limits are shown in Table \ref{tab:zprime}.

\begin{table}[htbp]
    \centering
    \begin{tabular}{l|rr|rr|rr}
         & \multicolumn{2}{c|}{250 GeV} & \multicolumn{2}{c|}{500 GeV} & \multicolumn{2}{c}{1 TeV}\\
         Model & excl. & disc. & excl. & disc. & excl. & disc. \\ \hline
         SSM & 7.7 & 4.9 & 13 & 8.3 & 22 & 14 \\
         ALR & 9.4 & 5.9 & 16 & 10 & 25 & 18 \\
         $\chi$ & 7.0 & 4.4 & 12 & 7.7 & 21 & 13 \\
         $\psi$ & 3.7 & 2.3 & 6.3 & 4.0 & 11 & 6.7 \\
         $\eta$ & 4.1 & 2.6 & 7.2 & 4.6 & 12 & 7.8 
    \end{tabular}
    \caption{Projected limits on $Z^\prime$ bosons in standard models, from the study of $e^+e^- \to ff$. The values presented, given in TeV, are the 95\% exclusion limits and the 5$\sigma$ discovery limits for the successive stages of the ILC program up to 1 TeV.}
    \label{tab:zprime}
\end{table}

\section{Summary}

We updated the discovery and exclusion limit of $Z^\prime$ models at the ILC. Using angular distributions, $Z^\prime$ models of SSM, ALR and $E_6$ models can be probed at 3-9, 6-16, 11-25 TeV with $\sqrt{s} = 250$, 500 and 1000 GeV, respectively, showing an example of a large discovery/exclusion potential of $e^+e^-$ linear colliders for BSM searches with precise electroweak measurements. More detailed studies with more BSM models, more realistic studies including full-simulation studies in each center-of-mass energy are desired.





\bibliographystyle{JHEP}
\bibliography{ref}  


\end{document}